\documentclass{article}
\usepackage{algorithm}
\usepackage{algpseudocode}
\usepackage{amssymb}
\usepackage{amsmath}
\usepackage[
backend=biber,
style=alphabetic,
sorting=ynt
]{biblatex}
\usepackage{float}
\usepackage{geometry}
\usepackage{graphicx} 
\usepackage[colorlinks=true, linkcolor=blue, citecolor=blue, plainpages=false, pdfpagelabels=true, urlcolor=blue]{hyperref}
\graphicspath{ {./images/} }
\usepackage{hyperref}
\usepackage{makecell}

\usepackage{tabularx}
\usepackage{pgfplots}
\pgfplotsset{width=10cm,compat=1.9}

\title{Sisu: Decentralized Trustless Bridge For Full Ethereum Node}
\author{Billy Pham \\ {billy.pham@sisu.network}
\and Huy Le \\ {huy.le@sisu.network}}
\date{}
\begin{document}

\maketitle
\begin{abstract}
  In this paper, we present a detailed approach and implementation to prove Ethereum full node using recursive SNARK, distributed general GKR and Groth16. Our protocol's name is Sisu whose architecture is based on distributed Virgo in zkBridge\autocite{xie2022zkbridge} with some major improvements. Besides proving signature aggregation, we provide solutions to 2 hard problems in proving Ethereum full node: 1) any public key is valid under previous beacon state and 2) all public keys are pairwise distinct. Our solution does not require worker-to-worker communication and therefore reduce total worker-to-worker network traffic from terabyte of data to zero compared to zkBridge. This makes our approach suitable for emerging distributed prover markets and more decentralized compared to zkBridge. Our design is highly parallelable and capable of running on GPU for most parts.

\end{abstract}

\section{Introduction}
As the number of blockchains grows, there is an increasing need for an efficient and secured bridge between various chains. These bridges are responsible for billions of dollars of assets flowing through them and having a secured bridge is an utterly important problem.

With an exception of Bitcoin \autocite{nakamoto2009bitcoin}, most popular blockchains today support an execution program called smart contract. A bridge between 2 popular blockchains often has smart contracts deployed on 2 chains and a relayer that observes and initiates interactions with these contracts. To send assets or messages from chain $C1$ to chain $C2$, a user first interacts with the contract deployed on chain $C1$ in a transaction. After this transaction is included in $C1$, the relayer reads the transaction on $C1$ and updates the smart contract on chain $C2$. Depending on the type of assets or data, the smart contract on chain $C2$ might initialize data transfer for users on chain $C2$ or just update its internal state.

\begin{figure}[H]
    \centering
    \includegraphics[scale=0.4]{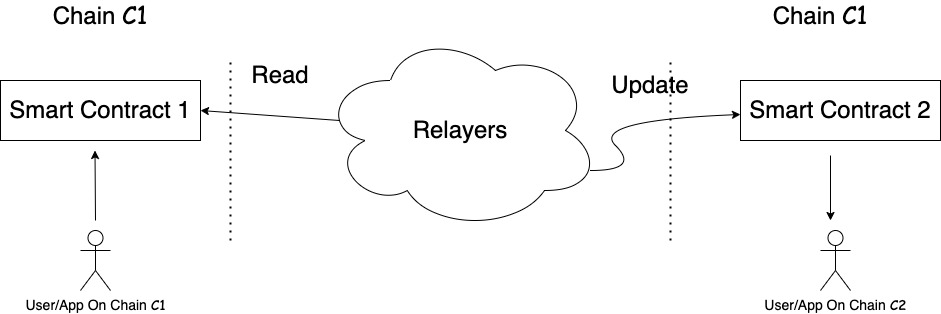}
    \caption{A common flow of how assets or data are transferred from chain A to chain B.}
    \label{fig:bridge_overview}
\end{figure}

Building a secured cross-chain bridge is a challenging task as each chain has its own protocol and working mechanism. In addition, who can update the smart contract on the destination chain decides the level of security of the bridge. There are a few types of bridges that have been built in practice, each depending on different cryptography assumptions: atomic swapping, threshold signature scheme, optimistic rollups and zero knowledge.

Though there are efforts to do research and build cross-chain bridges, many of them are known to be either insecured due to high centralization (threshold signature scheme) or having limited features to do general cross chain communication (atomic swap) or taking too long to withdraw assets (optimistic rollups). To date, billions of dollars have been lost through various bridge hacks \autocite{wormhole,rekt,harmonyhack}.

Recent advances in SNARK \autocite{chen2023review} and STARK \autocite{cryptoeprint:2018/046} research make building cross chain bridges using zero knowledge proof (ZKP) to be viable. Bridges that are built on ZKP are trustless, capable of supporting a wide range of applications and generally allowing users to quickly withdraw their assets. The main drawback of bridges based on ZKP, like many other ZKP applications, is the heavy computation that makes it difficult to run on a single machine or even on a small to medium cluster of machines within a small time frame.

The focus of this paper is on proving Ethereum full node using ZKP. Ethereum by far is the most popular blockchain that supports smart contracts. Any modern blockchain is very likely connected to the Ethereum network due to its popularity. We analyze why proving Ethereum full node is hard and how Sisu protocol addresses these challenges. The Ethereum version we refer to in this article is the version that uses Proof-of-Stake or Ethereum 2. This is different from the previous Ethereum version that runs on Proof-of-Work or Ethereum 1.

\subsection{The problems}

At the time of writing, Ethereum mainnet has about one million validator nodes \autocite{beaconscan}. For simplicity, we assume that there are one million active validators in Ethereum and each slot has 32,000 validators throughout this article.

To understand why proving Ethereum full node is hard, one needs to understand its underlying consensus protocol. Each of these one million validators is required to attest to exactly one Ethereum block in one epoch. Each epoch has 32 slots and each slot might contain at most one Ethereum block. There are roughly 32,000 validators producing 32,000 signatures in one block time frame. These signatures are aggregated and then included into the Merkle tree of the block.

To prove that one block is valid with regard to current beacon state, here is a list of necessary but not exhaustive tasks a prover needs to prove:
\begin{enumerate}
\item All these 32,000's signatures should be aggregated (in different committees) and valid under the new block hash.
\item Each of these signatures belongs to an active validator under the previous beacon state.
\item Each validator index must abide by an RANDAO algorithm which randomly shuffles and selects a set of validators for each epoch.
\end{enumerate}

The challenge of proving Ethereum full node is generating proof for 32,000 nodes. The second and third tasks require proving multiple SHA256 hashes. To prove the second task for one validator, a prover needs to show that the validator's public key is part of a Merkle path whose root is the previous beacon state hash. Ethereum consensus uses SHA256 for hashing and the current best SHA256 proof requires about 27,000 constraints \autocite{sasson2014zerocash}. Each path length from validator info leaf to the root is between 50 and 60. We are looking at 1,350,000 constraints per Merkle path per validator. Multiplying this number by 32,000 yields 43,2 billion constraints.

Proving the third task is even more challenging as each RANDAO to get the proposer index for a block requires multiple SHA256 hashes. We estimate that proving all validator indexes are correct under RANDAO requires between 100 billion to 1 trillion gates. Even the lower range estimation would exceed the computational capacity of the current best proving system. Note that these are not the complete list of problems we have to prove for Ethereum full node but the ones we want to highlight.

\subsection{Related Work}

zkBridge\autocite{xie2022zkbridge} is one of the first attempts to build a zero knowledge based bridge. It provides a high level approach in building a zero knowledge bridge using an algorithm called deVirgo which is a distributed version of Virgo protocol\autocite{zhang2020transparent}. However, zkBridge does not address in detail some of the hardest issues in proving Ethereum full node that we discuss in the previous section. It also requires a gigantic amount of network traffic for its worker-to-worker communication. It is estimated that proving signatures for 32,000 nodes in one block might generate more than 32 Terabyte of network traffic for worker-to-worker communication.

Some other projects \autocite{telepathy} build bridges based on zero knowledge proof but for Ethereum sync committee instead of full nodes. Ethereum protocol has a sync committee of 512 validators selected every sync committee timeframe or about a day. Generating proof for the sync committee is faster than doing it for full node because only one proof is needed per day. The security level of this approach is much lower compared to proving full nodes as 512 is less than 0.1\% of total validators in Ethereum. We believe this security level is not strong enough in the events when the stake is high.

\subsection{Our Contribution}
Our contributions of Sisu protocol in this paper is as follow:
\begin{itemize}
\item We introduce an algorithm to prove that the majority (more than 2/3) of validators in Ethereum network in one epoch are validators under the beacon state in the previous epoch and they are all pairwise different. A naive approach compares all pairs of validators public key and runs in $O(n^{2})$. We propose an algorithm that runs on $O(n)$ in circuit by leveraging special properties of the validator set.

We argue that by proving the uniqueness of the majority of validators in one epoch we can replace the RANDAO algorithm of Ethereum consensus with our new method. To verify one block, the RANDAO algorithm defines a specific set of validators while Sisu protocol requires any set of valid validators. However, within one epoch the set of validators of RANDAO is the same as the set of validators in Sisu protocol. This set is the set of active validators in Ethereum network. The cost of attacking Sisu protocol for this part is equivalent to the cost of manipulating the majority of Ethereum networks.

\item We propose a new approach that requires no worker to worker and minimal traffic between worker and master in our distributed system. This allows Sisu to scale better compared to the design in \autocite{zhang2020transparent}.

\item Our protocol is built on a distributed GKR system using a general circuit. Compared to a layered circuit, the general circuit reduces the number of layers in the circuit and achieves 3-4 times speed up in performance.

\item Our circuit uses a new type of gate called accumulation gates that allows more than 2 input gates connecting to the same output gate without modifying the GKR evaluation function. Though the number of evaluations remains the same, accumulation gate significantly reduces the number of layers and the number of gates per layer.

\item We incorporate GPU hardware acceleration in our implementation with multiple optimization strategies. We first implement Sisu protocol in CPU, identify the part that could run in parallel and gradually port them to GPU.
\end{itemize}
Sisu protocol is designed toward decentralized proving. We make a deliberate trade-off of having decentralization at the expense of increasing the number of constraints. However, with the emerging prover markets we expect ZK applications in the future could run on hundreds and even thousands of GPUs. Any protocol that can be divided into independent provers could take advantage of distributed proving systems.

\section{Background}

\subsection{Zero Knowledge Proof}

An argument system for an NP relationship $\mathcal{R}$ is a protocol between a computationally-bounded prover $\mathcal{P}$ and a verifier $\mathcal{V}$. At the end of the protocol, $\mathcal{V}$ is convinced by $\mathcal{P}$ that there exists a witness $w$ such that ($x$; $w$) $\in R$ for some input $x$. We use $\mathcal{G}$ to represent the generation phase of the public parameters pp. Formally, consider the definition below, where we assume $R$ is known to $\mathcal{P}$ and $\mathcal{V}$.

Let $\mathcal{R}$ be an NP relation. A tuple of algorithm ($\mathcal{G}$, $\mathcal{P}$, $\mathcal{V}$) is a zero-knowledge argument of knowledge for $\mathcal{R}$ if the following holds.

\begin{itemize}
    \item \textbf{Correctness}. For every $pp$ output by $\mathcal{G}$($1^\lambda$) and ($x$, $w$) $\in$ R and $\pi \leftarrow$ $\mathcal{P}$($x$, $w$, $pp$),
    \[ Pr[\mathcal{V}(x,\pi,pp)=1] = 1\]
    \item \textbf{Soundness}. For any PPT prover $\mathcal{P}$, there exists a PPT extractor $\varepsilon$ such that for every $pp$ output by $\mathcal{G}(1^\lambda)$ and any x, the following probability is \textbf{negl}($\lambda$):
    \[ Pr[\mathcal{V}(x,\pi,pp)=1 \wedge (x,w) \notin R | w\leftarrow\varepsilon(pp,x)] \]
    \item \textbf{Zero Knowledge}. There exists a PPT simulator $\mathcal{S}$ such taht for any PPT algorithm $V^{\ast}$, auxiliary input $z \in \{0,1\}^{\ast}, (x,w) \in \mathcal{R}$, $pp$ output by $\mathcal{G}(1^{\lambda})$, it holds that
    \[ View(\mathcal{V}^{\ast}(pp,x))\approx\mathcal{S}^{\mathcal{V}^{\ast}}(x,z) \]
\end{itemize}

The tuple ($\mathcal{G}, \mathcal{P}, \mathcal{V}$) is a succinct argument system if the running time of $\mathcal{V}^{\ast}$ and the total communication between $\mathcal{P}$ and $\mathcal{V}$ (proof size) are poly($\lambda$, $\vert x \vert$, log$\vert w \vert$).

\subsection{GKR Protocol}

GKR is a protocol proposed in \autocite{goldwasser2008delegating} where the soundness of the output layer \(L_0\) can be reduced to previous layer and all the way to input layer \(L_D\). At each layer \(L_i\), both prover and verifier run sumcheck protocol to reduce the soundness to layer \(L_{i+1}\). This process keeps going until the layer input. At the end of the protocol, the verifier checks the soudness from the given input x.

We assume that a layer i-th has \(S_i\)=\(2^{s_i}\) gates (we can add more padding gate to the layer if this is not true) and these gates are indexed using \(s_i\) bit numbers. We define a function \(V_i\): \(\{0,1\}^{s_i}\) $\rightarrow$ $\mathbb{F}$ that maps a binary string of length \(s_i\) of \{0,1\} and returns the output of the gate. The output value of gate \(g\) is derived from the output values of gates from previous layer:

\[ {V_i}(g) = \sum_{x,y \in \{0,1\}^{2s_{i+1}}} add_i(g,x,y)(V_{i+1}(x)+V_{i+1}(y))+mult_i(g,x,y)(V_{i+1}(x).V_{i+1}(y))\] where \(add_i\) and \(mult_i\) returns \(1\) if there are connection between output gate \(g\) in layer \(i\) to layer \((x,y)\) in layer \(i+1\). \(V_i\) could be extended to its \textit{multilinear extension} form $\tilde{V_i}$:

\[ \tilde{V_i}(g) = \sum_{x,y \in \{0,1\}^{2s_{i+1}}} \widetilde{add_i}(g,x,y)(V_{i+1}(x)+V_{i+1}(y))+\widetilde{mult_i}(g,x,y)(V_{i+1}(x).V_{i+1}(y))\] where $\widetilde{add_i}$ and $\widetilde{mult_i}$ are multilinear extensions of \(add_i\) and \(mult_i\) respectively.

The novel contribution of GKR protocol in zero knowledge proof protocol is that the prover runs in polynomial time of circuit size and the verifier could run in \(sublinear\) time. Specifically if the circuit has \textit{highly regular wiring pattern} then $\widetilde{add_i}$ and $\widetilde{mult_i}$ can be evaluated in $\mathcal{O}(logS_i)$. The total running for $\mathcal{V}$ in that case is $\mathcal{O}(n+dlogS)$ with \(d\) is the depth of the circuit and \(n\) is the input size.

The GKR protocol can be sped up further if the computation of the circuit could be divided into independent sub-circuit computation before final aggregation \autocite{thalerbook2022} . These 2 optimizations of GKR are exploited to speed up proving in Sisu protocol.

\subsection{Libra protocol}
In \autocite{thaler2013time},  Thaler proposed a linear-time algorithm for the prover of the sumcheck protocol on a multilinear function $f$ on $l$ variables (the algorithm runs in $O(2^l)$). Based on Thaler's proposal, Libra \autocite{xie2019libra} introduced a linear time sumcheck protocol on product of multilinear functions
\begin{equation}
    f(X)=f_1(X)*f_2(X)=f_1(x_1, x_2, ..., x_l) * f_2(x_1, x_2, ..., x_l)
\end{equation}

The protocol uses two bookeeping tables to generate the univariate polynomial in each round. These bookeeping tables are essentially the evaluations of the function $f_1(x_1, x_2, ..., x_l)$ and $f_2(x_1, x_2, ..., x_l)$ where $x_1, x_2, ..., x_l$ run on boolean hypercube. At round $i$ ($i=1 \ldots l $), each bookeeping table has a size of $2^{l-i-1}$. The complexity of the algorithm producing the univariate polynomial and updating tables at each round is linear with the size of the table.

Libra also proposes an algorithm to run sumcheck on $\widetilde{mult_i}(g,x,y)(V_{i+1}(x).V_{i+1}(y))$ in linear time by splitting this function into two phases: the first one runs the sumcheck protocol on only variable $x$, meanwhile the second one runs the sumcheck on variable $y$. This helps Libra achieve the linear prover time in GKR protocol.

\subsection{Virgo protocol}
A naive implementation of GKR needs the verifier to evaluate $\tilde{V_D}$ in the last round. \autocite{zhang2020transparent} addresses the issue of GKR in Virgo, a transparent SNARK protocol on a structured circuit with logarithmic proof size and verification time. In Virgo, instead of having a verifier to evaluate $\tilde{V_D}$, the prover is asked to commit to $\tilde{V_D}$ before running the GKR protocol using Verifiable Polynomial Delegation (VPD). At the end of GKR protocol, the verifier asks the prover to open $\tilde{V_D}$ at 2 random points and validates them using VPD.Verify. With this approach, the verifier does not need to know the value of \(witness\) while holding GKR soundness because of the soundness of VPD.

The VPD algorithm in Virgo does not require trusted setup and is inspired by univariate sumcheck protocol described in \autocite{ben2019aurora}. The verifier in VPD is optimized to reduce its complexity from $O(n)$ to $O(log^2n)$ where $n$ is the input size. A recursive proof system that uses Virgo can benefit from this reduction as the prover of a proof system after Virgo has to simulate Virgo's verifier in circuit to verify Virgo's proof.

\subsection{Recursive SNARK and Computation Compression}
Recursive SNARK was first introduced in \autocite{bitansky2013recursive} and subsequently used in many protocols such as Halo \autocite{bowe2019recursive} to improve verification time as well as the proof size. A recursive SNARK is a proof system that contains many other SNARK inside. In many cases like Nova \autocite{kothapalli2022nova} this is a list of consecutive proofs connected to each other. In some other cases like zkTree \autocite{deng2023zktree}, the nested proofs have tree structure. In a recursive SNARK, a proof is generated for each nested proof system or each layer of computation or recursion. These proofs are then recursively composed to generate proof for the entire system.

Each SNARK protocol has its own strength and weakness in terms of proof length, prover time, verifier time or various degree of security. For example Groth16 \autocite{groth2016size} has short proof and fast verification time but long proving time (\(O(NlogN))\) and requires trusted setup. Other framework such as Orion \autocite{xie2022orion} has faster prover time but longer proof size. Combining multiple proof systems in a recursive proof can leverage the strength of multiple proof systems.

Consider a recursive SNARK system with large input size that uses Orion as the first proof followed by Groth16. The output proof of Orion is used as an input for Groth16 which in turn produces a final short proof. Since the proof output of Orion is usually much smaller compared to its input size, applying the (\(O(NlogN))\) of Groth16 on the output reduces the running time for the prover compared with applying Groth16 directly to the original input.

\subsection{zkBridge protocol}
zkBridge is a recursive proof system with 2 phases: a distributed version of Virgo called \(deVirgo\) followed by Groth16 to reduce the proof size. Since proving a new Ethereum block involves proving multiple identical tasks such as signature verification, this opens the opportunity for running the prover in a distributed manner on multiple machines.

Specifically, there are \(N\) machines labeled from $\mathcal{P}_0$ to $\mathcal{P}_{N-1}$ with $\mathcal{P}_0$ as the master node. \(deVirgo\) does not naively apply Virgo to each identical sub-circuit and then aggregate these proofs as the final proof size is proportional to the \(N\) and this could be expensive to run in second phase Groth16. Instead, the protocol uses a new distributed version of GKR and distributed polynomial commitment. In this scheme, the master node $\mathcal{P}_0$ combines messages from all machines before sending the aggregated message to the verifier. The proof size is reduced by a factor of \(N\) compared to the naive approach. The downside of this scheme is the increased total network communication which can quadratically be proportional to the number of machines. zkBridge requires a very centralized approach as the heavy network traffic makes distributed provers an infeasible option.

\subsection{General Arithmetic Circuit}

\autocite{zhang2021doubly} introduces a protocol that generalizes layered GKR circuits to arbitrary arithmetic circuits while preserving the performance of the prover. The difference between these 2 types of circuits is layered circuits gates in layer \(i)\) can only connect to input gates in layer \({i+1}\) while in general circuits these gates can connect to gates belonging to any previous layer \(j\) as long as at least one input gate is in the layer \(i+1\). General circuit reduces the number of gates used in the circuit since gates in layer \(i\) can use input gates in arbitrary layer \(j\) without forwarding it to layer \(i+1\).

Another major difference between general and layered circuits is how input gates are indexed. In a layered circuit, gates are indexed using their absolute position in the layer. In general circuits, for any pair of layers \(i\) and \(j\) where there is at least one wiring only used gates are indexed and these indexes are different from their absolute position in the layer. Because the indexes of used gates are only a subset of gates used in layer \(j\) we can preserve the performance of the verifier.

\subsection{Distributed Prover Market}
In spite of many recent improvements in SNARK performance \autocite{chen2023hyperplonk,bunz2023protostar}, generating proof for many ZKP applications is still resource-intensive work. Most current implementation relies on centralized proving system which might be vulnerable for censorship and single point failure. A distributed prover market can provide large scale proving system as a service, resistant to censorship and cost saving utilities for users through competitive price auction.

There are a number of projects that actively working on fee mechanism for ZK proofs such as \textit{Proo$\varphi$} \autocite{wang2024mechanism}, Nil \autocite{nilprotocol}, Gevulot \autocite{gevulot}, etc. These markets provide a common ground for users and proof service providers to transact and generate proofs with some fee. Distributed prover market is a good fit for our protocol which is designed to run independently for most parts. We leave the choice of using what prover markets for Sisu implementation to the operator and future work.

\section{Sisu Protocol}
Although zkBridge proposes a practical approach for proving circuits with a large number of parallel sub-circuit, a straightforward implementation of zkBrigde might not be sufficient for proving Ethereum full node as it might require between a few hundred billions to a trillion circuit gates. Moreover, zkBridge uses a massive amount of network communication that creates a significant pressure on networking bandwidth and processing time even when running within a data center.

Sisu protocol is inspired by zkBridge and other SNARK proof systems \autocite{zhang2020transparent} with many major improvements to address the issues in zkBridge. While Virgo uses a layered circuit for its proving scheme Sisu uses a general arithmetic circuit. General arithmetic circuit reduces the number of gates and layers in the circuit and has 8-10x speed up compared to a layered circuit. Secondly, we propose a new way to distribute work between worker machines using cluster shared memory pool that does not require worker to worker communication. Network traffic between worker and master nodes is minimal. Sisu addresses networking issues in zkBridge and yields better scalability in terms of networking traffic.

We provide a new method to check that validator's public keys are distinct in one epoch. A naive implementation would do comparison for all pairs of public keys and with \(O(N^2)\) running time. This implementation is infeasible to apply for one million keys and above. Sisu leverages special properties of the Ethereum block to reduce verifier's running time to \(O(N)\).

\subsection{Overview}

In the Sisu protocol, proof is generated for one block at a time. The prover's public inputs are: new block hash, previous beacon state hash, public key index hash and demonstrates that the new block is an immediate valid block after a previous beacon state. The verifier is an updater smart contract deployed on-chain. The prover generates a proof and submits it to the updater contract. If verification succeeds, the updater contract adds the new block with its new beacon state to its list of accepted blocks. Applications who want to check if the status of an object in a remote chain in a block can send a query of Merkle path to the updater contract.

The prover in Sisu protocol uses a recursive proof argument with 2 phases: the first phase consists of multiple parallel sub-circuit computation and the second phase uses Groth16 to verify the proof from the first phase and reduces the final proof size. These phases can have overlapping execution on different blocks. For example, while a machine is applying phase 2 for block 0 another machine can start processing phase 1 of block 1.

\begin{figure}[H]
    \centering
    \includegraphics[scale=0.6]{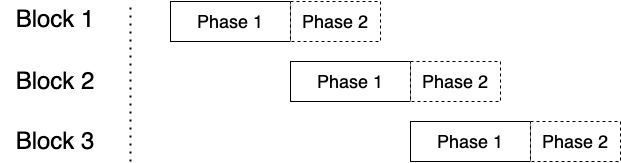}
    \caption{Overlapping execution of phase 1 and phase 2 for different blocks}
    \label{fig:phase1_phase2}
\end{figure}

The total proof time of the whole protocol is the greater proof time of phase 1 or phase 2. We set a target of optimizing this number to be smaller than one Ethereum block time or 12 second. Phase 1 uses distributed general circuits (DGC) for efficient prover. Phase 2 uses Groth16 to reduce the proof size to constant size.

\subsection{Distributed General Circuit}

In this section, we describe the architecture for a distributed general circuit which runs on \(N\) machines, each handles the computation of one sub-circuit. On a high level idea, our DGC protocol uses distributed sumcheck and distributed polynomial commitment similar to what are described in zkBridge. However, the distributed sumcheck in zkBridge applies for the general function \(f\) while in our case \(f\) is a product of multiple linear functions. There are a few differences in how our function \(f\) is calculated.

\subsubsection{Distributed multilinear product sumcheck}
\textbf{Background: distributed sumcheck for general function \(f\)}.  Suppose the prover has \(N\) machines $\mathcal{P}_0$, $\mathcal{P}_1$,..., $\mathcal{P}_{N-1}$. Each machine $\mathcal{P}_i$ holds a polynomial \(f^{(i)}\): $\mathbb{F}^{l-n}\to{\mathbb{F}}$ such that \(f^{(i)}(x)\) = \(f(x[1:l-n],\textbf{i})\) with \textbf{i} is a binary representation of index i. deVirgo distributed sumcheck proceeds in \(l\) rounds:

\begin{enumerate}
    \item In the \(j^{th}\) round where 1 $\leq$ j $\leq l-n$, each $\mathcal{P}_i$ sends $\mathcal{P}_0$ a univariate polynomial
    \[ f^{(i)}_j(x_j) = \sum_{\textbf{b} \in \{0,1\}^{l-n-j}} f^{(i)}(\textbf{r}[1:j-1],x_j,\textbf{b})\]
    After receiving all univariate polynomials from $\mathcal{P}_1$,..., $\mathcal{P}_{N-1}$, $\mathcal{P}_0$ computes:
    \[ f_j(x_j) = \sum_{i=0}^{N-1} f^{(i)}_j(x_j)\]
    and then sends $f_j(x_j)$ to the Verifier to check $f_{j-1}=f_j(0)+f_j(1)$, and sends a random challenge $r_j \in \mathbb{F}$ to $\mathcal{P}_0$. $\mathcal{P}_0$ relays \(r_j\) to $\mathcal{P}_1$,...,$\mathcal{P}_{N-1}$.
    \item In the $j=(l-n)^{th}$ round, each $\mathcal{P}_i$ computes \(f^{(i)}(r[1:j])\) and sends it to $\mathcal{P}_0$. $\mathcal{P}_0$ then constructs a multi-linear polynomial \(f'\): $\mathbb{F}^{l-n}\to{\mathbb{F}}$ such that $f'(i)=f^{(i)}(r[1:j])$ for 0 $\leq$ i $< N$.
    \item in the $j^{th}$ round, where $l-n<j\leq l$, $\mathcal{P}_0$ continues to run the sumcheck protocol on $f'$ with the verifier.
\end{enumerate}

\textbf{Distributed multilinear product sumcheck}.
In our protocol, instead of calculating an arbitrary function \(f\) we need to calculate the sumcheck of this kind of function (where $f_i$ are multilinear polynomials):

\begin{align}
 f =  \nonumber &\sum_{x \in \{0,1\}^l}f_0(x)f_1(x) \\ \nonumber
 + &\sum_{x \in \{0,1\}^l}f_0(x)f_2(x) \\ \nonumber
 + &\ldots \\
 + &\sum_{x \in \{0,1\}^l}f_0(x)f_i(x) \\ \nonumber
 + &\ldots \\ \nonumber
\end{align}

If we apply the algorithm of distributed sumcheck for general polynomial directly, the proving time for (1) is not linear. The sumcheck algorithm of general arithmetic circuit is described in \autocite{zhang2021doubly} and Libra protocol \autocite{xie2019libra} where the prover needs to maintain a list of bookeeping tables of each $f_i$ to run the sumcheck protocol for product of multilinear polynomial in the linear time. At each round, the prover uses these bookeeping tables to produce the univariate polynomial, then update them to prepare for the next round, these two operations are linear time in size of polynomial. In the distributed scenario, at $(l-n)^{th}$ round, instead of send $f^{(i)}(r[1:j])$,  $\mathcal{P}_i$ needs to send value of all $f_k^{(i)}(r[1:j])$ (they are also the last value in the each bookeeping table) to the master. The master collects these values, rebuilds bookeeping tables (each table has size of $N$), then continues to run the multilinear product sumcheck protocol based on these tables. This allows the master runs sumcheck protocol in the linear time.

\subsubsection{Distributed polynomial commitment}
Protocol 6 in zkBridge presents an algorithm to do polynomial commitment across N machines for a function $f$ for a point $\textbf{r} \in \mathbb{F}^{l}$. The formula to calculate $f(\textbf{r})$ is:
\[ f(\textbf{r}) = \sum_{i=0}^{N-1} \tilde{\beta}(\textbf{r}[l-n+1:l],\textbf{i})f^{(i)}(\textbf{r}[1:l-n]) \]
where $\tilde{\beta}(\textbf{x},\textbf{y})=\prod_{i=1}((1-x_i)(1-y_i)+x_iy_i)$ is an identity function. A naive approach let $\mathcal{P}_i$ commit $f^{(i)}$ separately and creates $N$ proofs for verifier. The issue with this approach is the size of the proof as it is proportional to $N$.

zkBridge addresses this issue by assembling all proofs from $N$ machines first before sending them to the verifier. Each worker machine $\mathcal{P}_i$ exchanges data with other worker machines immediately after calculating its portion $f^{(i)}$. After receiving all portions from other machines, $\mathcal{P}_i$ then creates a commitment $com_{h^{(i)}}$ and sends it to $\mathcal{P}_0$. $\mathcal{P}_i$ assembles all the commitments and sends the final message to $\mathcal{V}$. Though this method reduces the proof size, it requires $O(N) * data\_size $ network communication per machine and $O(N^2) * data\_size $ total network communication cost. In signature proof alone generation, zkBridge produces approximately 1GB of data per machine. Doing this for at the scale of Ethereum full node would require multi terabyte of data in network traffic. Even when these machines in the same data center, this not only creates pressure on network infrastructure but also increase each machine's processing time due to idle waiting and synchronization.

The naive approach and zkBridge's approach is a trade off between computation and network traffic. The former has zero worker-worker communication (though it still has some worker-master traffic) but it has big proof size and increases $\mathcal{V}$ running time. On the other hand, the zkBridge's approach has minimal proof size at the cost of massive network traffic. We can have solutions that stay in between these two extremes: a sublinear proof size and some worker-worker traffic. Sisu improves this even furthur by having sublinear proof size (compared to naive approach) and zero worker-worker traffic.

In step 2 and 3 in protocol 6 of distributed PC, each $\mathcal{P}_i$ exchanges data with other machines before creating its own commitment. Let $s_j$ is a list of of $j^{th}$ element of all machines. In other words, $s_j$ = [$f^{(i)}_j$  for $j$ in 0..$N$-1]. We need to hash this array to create the commitment for $\mathcal{P}_i$. This array could be hashed using Poseidon \autocite{grassi2021poseidon} or any array friendly hash.

\subsubsection{Distributed polynomial commitment with shared cluster mempool}

We first divide $N$ machines into $K$ clusters, each with $M$=$N$/$K$ machines and a leader. Each cluster creates its own commitment and sends the proof to $\mathcal{V}$. Compared to zkBridge's approach, our proof size grows by a factor of K. In addition, creating a cluster proof still requires $O(M^2)*data\_size$ work-worker network communication. To eliminate this network communication completely, we let each machine write its own data to its designated area in a shared array and then notify the cluster leader. The leader waits for all machines to finish writing before assembling commitments to create the final proof.

\begin{figure}[H]
    \centering
    \includegraphics[scale=0.4]{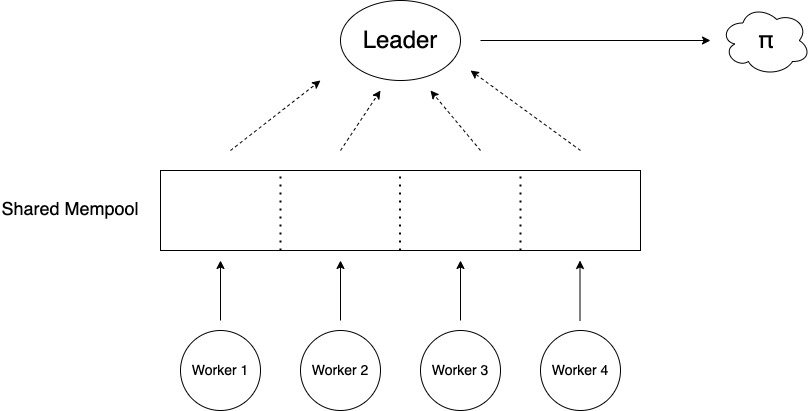}
    \caption{Each worker independently writes to its designated area in the mempool. The cluster leader later assembles all the pieces of data to create final proof.}
    \label{fig:shared_mempool}
\end{figure}

A good value of $K$ varies between $log(N)$ and $\sqrt{N}$. The exact choice of $K$ depends on how much computation when we want to shift to $\mathcal{V}$ and the type and availability of shared mempool. A large mempool can accommodate more workers per cluster and hence reduces the value of $K$. This approach is a trade off of having separation of machine execution at the expense of increased constraints count. This might need more hardware to execute but this model fits well with distributed prover markets described in previous sections.

\subsection{Circuit designs}

\subsubsection{Public key aggregation}

The Gasper protocol \autocite{buterin2020combining} and Capella fork consensus \autocite{consensus2024} describe how a validator attests a block. In \autocite{buterin2020combining} a committee is defined as a set of all validators in one slot while the \autocite{consensus2024} defines a committee is a fraction of validators in a slot. We use the latter definition of committee in our protocol. In a slot, 32,000 validators are divided up to 64 committees of approximately equal size using a RANDAO algorithm. Validators in one committee provide attestation to the same block in a previous slot. Signatures from these validators are aggregated to create one signature of the whole group.

One nice property of the BLS12-381 curve is that verifying signature aggregation of \(N\) validators is equivalent to aggregating public keys of these validators first and then verifying this aggregated public key with the aggregated signature. In other words, we can do public key aggregation instead of signature aggregation and then verifying the aggregated public key with an aggregated signature. Verifying a pair (public key, signature) is expensive in circuit while public key aggregation is relatively cheap.

The aggregation of 2 public keys is simply an addition of point representation of these keys on the BLS12-381 curve. The aggregation of all validators' public keys in a committee is the sum of their respective points on the curve. The first aggregation subcircuit in Sisu proves an addition of validators' public key \(v1\) and \(v2\) and outputs a new public key \(v12\). The second subcircuit proves an addition of \(v12\) and \(v3\) key and output key \(v123\), the third subcircuit proves an addition of \(v123\) and \(v4\) and so on.

\begin{figure}[H]
    \centering
    \includegraphics[scale=0.4]{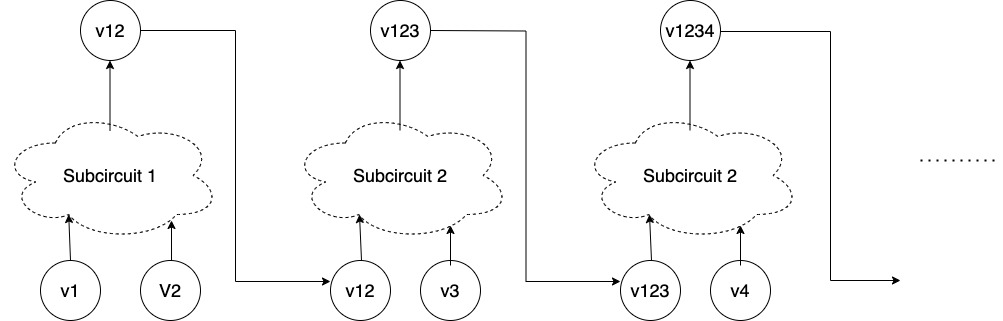}
    \caption{An overview of public key aggregation. Besides verifying each subcircuit's proof, the verifier also checks if the output of one circuit is in the input of the next circuit.}
    \label{fig:pubkey-aggregation}
\end{figure}

Addition on a curve is \textit{associative} which means the order of numbers is not important. Instead of placing each validator's public key in the same order in their committee, the prover sorts them by the validator indexes. This sorted list is used when proving uniqueness of each validator without changing the final aggregation public key.

\subsubsection{Verifying validator's authenticity}

Proving public key aggregation is not sufficient as a malicious prover can make up new public keys and create a new block hash that contains the Merkle path of these public keys. The prover also needs to demonstrate that these public keys are valid under current beacon state.

In native program execution, one can cache the list of all validators in memory and use this list to run the RANDAO algorithm to calculate the validator indexes and get their information in a slot. When proving in a circuit, we do not have access to this cached list and cannot use the same method to verify aggregated signatures. Fortunately, a validator's public key is part of a beacon state hash and a prover can prove the validity of a validator by providing its Merkle path to the beacon state root.

Ethereum consensus hashes all objects in the consensus layer using Simple Serialize (SSZ) hash. SSZ is designed to be deterministic and efficient when creating a Merkle tree. All validators, including active and non-active ones, are included in the Merkle tree. Ethereum limits the number of validators to be 4 billion. Instead of changing the validator list size whenever a new one joins the network, Ethereum creates all 4 billion slots for all possible validators and assigns empty slots with validator index 0.

To show that a validator is active under the current beacon state, we need to provide the Merkle path from the public key leaf to the beacon state root. This translates to $d$ SHA256 circuits to prove a single Merkle path where $d$ is the depth from the root to the public key node. In practice, the value of $d$ is between 50 and 60. Fortunately, only a small fraction of 4 billion validators are activated validators. This means that for the subtree that represents the validator group, most of the hash values are hashes of 0 and remain unchanged. We can take advantage of this fact to shorten the Merkle path length that we have to prove.

\subsection{Proving public pairwise distinct}

Besides showing that all validators' public keys are valid under a beacon state root, $\mathcal{P}$ is required to demonstrate that all these public keys are pairwise different. If this requirement is not present, a malicious prover with control of at least one valid validator can create multiple replicas of its public key in the circuit to produce counterfeit blocks. By requiring $\mathcal{P}$ to show that all the public keys are pairwise different, we make it extremely hard for a malicious prover to forge a new block unless it has the control of a majority of validators in a slot.

Each Merkle path can be represented as a list of hashes from a leaf and all sibling (left or right) hashes to the root. From this path, a validator can calculate the index of the validator. In Sisu protocol, the prover is required to provide a list of Merkle paths sorted by validator indexes. Checking if a list is sorted or not and checking if a sorted list has distinct elements are trivial and both tasks can be done in linear time.

The cost of forging a block is the cost of forging the majority ($\frac{2}{3}$) of validator voting power in a block. Since each validator is required to deposit at least 32 ETH to be activated, this attack requires 682666 ETH to execute. At the time of writing, one ETH's price hovers around \$3000 and this attack costs about 2 billion USD. Note that this does not have the same level of security as RANDAO since RANDAO requires the attacker to control a specific set of randomized validators while in our protocol the attacker can control $\textit any$ set of validators.

If the cost of forging a block is not high enough, our protocol makes it more difficult for malicious provers by requiring the majority of $\textit all$ validators must produce signatures in one epoch and are pairwise distinct. This raises a challenging question as our protocol processes one block at a time and the verifier has no knowledge of the previous block. Before answering this question, we introduce a new hash function to hash all validators' indexes.

\subsubsection{Associative hash function}
An associative hash function is a function that hashes a list of numbers to the same result regardless of the order of numbers in the list. In other words, swapping any 2 numbers in the list does not change the hash result. Its formal representation is as follow:

\[ \mathcal{AH}([e_1, e_2, ... , e_n]) = \mathcal{AH}(PERM([e_1, e_2, ... , e_n])) \]
where \(PERM[a]\) is a function that returns a permutation of array $a$. Here are a few examples of associative hash function:

\begin{itemize}
    \item $\mathcal{AH}([e_1, ... , e_n]) = e_1 + e_2 ... + e_n$
    \item $\mathcal{AH}([e_1, ... , e_n]) = e_1 * e_2 ... *e_n$
    \item $\mathcal{AH}([e_1, ... , e_n]) = (e_1+1)^3 + (e_2+1)^e + ... + (e_n+1)^3$
\end{itemize}
One important constraint in our proof is that all indexes must be smaller than $n$ since Ethereum has $n$ active validator. In other words, a malicious prover not only has to find a set of numbers that hashes to the same value as hashing the original list but also makes sure that these numbers must be smaller than $n$. We also simplify our associative hash function to make it have a form $\mathcal{AH}([e_1, ... , e_n]) = \mathcal{F}(e_1) + \mathcal{F}(e_2) + ... + \mathcal{F}(e_n)$ where $F$ is a function that applies on individual function.

\[\left\{
    \begin{array}
    {rcl}\mathcal{F}(e_1) + \mathcal{F}(e_2) + ... + \mathcal{F}(e_n) = \mathcal{H} \\
    0 \leq e_1 \leq n \\
    0 \leq e_2 \leq n \\
    ... \\
    0 \leq e_n \leq n \\
    \end{array}
\right.\]

We will demonstrate that given $n$ values $e_1$, $e_2$, ..., $e_n$, the probability that a malicious prover can find a second solution for the system of equations above is negligible. We first choose a function $\mathcal{F}$ that has an avalanche effect or any small change in the input can lead to large unpredictable output. Secondly, the probability to find a solution $e_i$ such that $0 \leq e_i \leq n$ is $n/p$ and the probability of finding all solutions for this system of equations is $({n/p})^n$. With $n$ is one million and $p$ often in the range $2^{64}$ to $2^{381}$, this probability is negligible.

The choice of our function $\mathcal{F}$ is provided in the algorithm \ref{alg:associative_hash_f} below for our field BN254. Other fields can derive another function $\mathcal{F}$ with similar design strategy.

\begin{algorithm}
  \caption{Associative Hash: hash one single element}\label{alg:associative_hash_f}

  \textbf{Input}: An element $e$ on finite field BN254. \\
  \textbf{Output}: Hash of the element $e$ on field BN254.

  \begin{algorithmic}
      \Function{F}{$e$}
          \State {$r$ $\gets$ $0$}

          \For{$i \gets 1$ to $3$}
            \State {$r$ $\gets$ $(r+e+4294967295)^3$}
          \EndFor

          \State \Return {$r$}
      \EndFunction
  \end{algorithmic}
\end{algorithm}

From a first glance, this function looks similar to Mimc function but with much less rounds. The number $4294967295$ is $2^{32}-1$ but could be replaced by other similar large number. The idea of adding this big number and then raise to low power is to make sure that any small change in input can lead to drastic change in output. For example, the difference between $1001^2$ and $1000^2$ is much bigger than the difference between $5^2$ and $4^2$. This gap is even wider when these numbers are raised to bigger power.

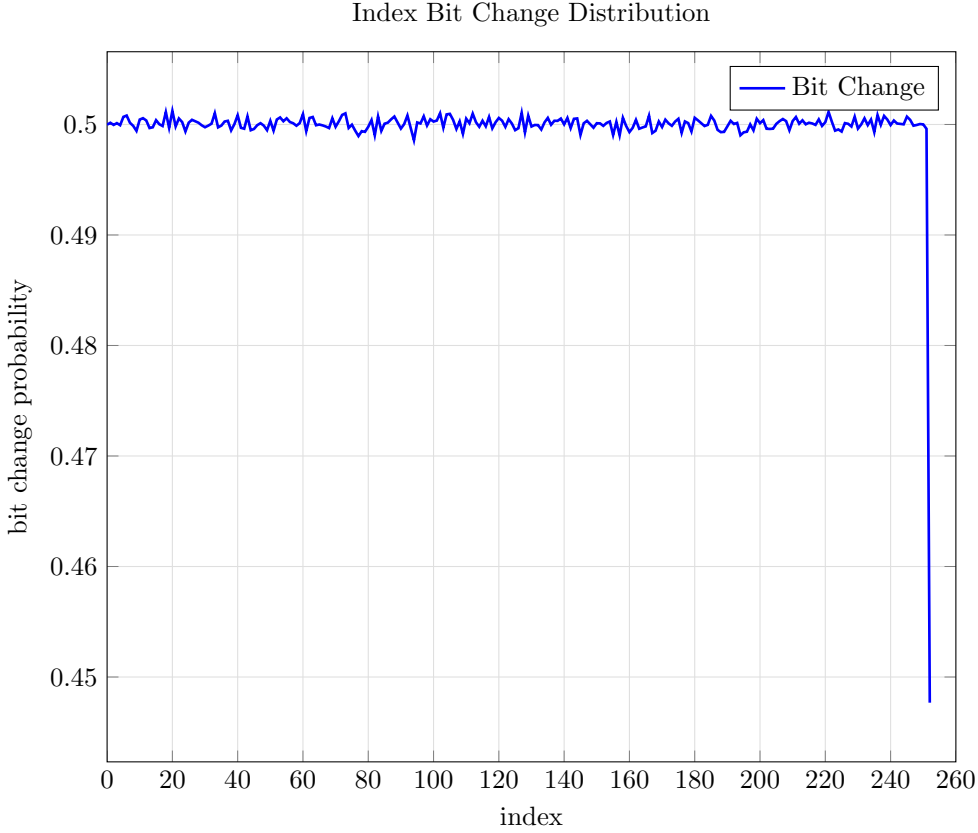
\begin{figure}[H]
  \centering
  \begin{tikzpicture}
    \begin{axis}[
      xlabel=index,
      ylabel=bit change probability,
      title={Index Bit Change Distribution},
      xmin=0,
      xmax=260,
      grid=both,
      minor grid style={gray!25},
      major grid style={gray!25},
      width=0.75\linewidth,
      no marks]
    \addplot[line width=1pt,solid,color=blue] %
      table[x=index,y=bit_change,col sep=comma]{index-bit_change.csv};
    \addlegendentry{Bit Change};
    \end{axis}
  \end{tikzpicture}

  \caption{Bit Change Probability of $\mathcal{F}(x+1)$ - $\mathcal{F}(x)$}
  \label{fig:bit-change-rate}
\end{figure}

One important difference between our hash function and Mimc is that we do not care about pre-image or key recovery attacks since all data is public. We only care about if this function is close to uniformly distributed and if one small change in input can lead to random change in output. We run an experiment to show that this function is close to having a uniform distribution. We calculate $\mathcal{F}(x+1)$ - $\mathcal{F}(x)$ for one million number from 1 to 1,000,000 and see how a small change in input can affect little endian output bits in figure \ref{fig:bit-change-rate}. We observe the the probability of any bit that is changed is within $0.5\pm0.1$. The drastic drop in probability in the last bit is caused by the fact that our prime number BN254 does not have equal distribution for the last bit.

We can expect the output of the function $\mathcal{F}$ is randomized enough for an attacker to manipulate the input. Any change in the input would lead to a new randomized output. It would be impossible for the malicious prover to find a new set of ($e_1$, $e_2$, ..., $e_n$) that sums up to a predefined value $\mathcal{H}$.

\subsubsection{Proving public keys pairwise distinct}

Assuming our associative hash function is hard to attack, we present an algorithm with running time \(O(N)\) for $\mathcal{V}$ and \(O(Nlog(N))\) for $\mathcal{P}$ to verify that all public keys are pairwise distinct. Given an unsorted array $A$, $\mathcal{V}$ outputs 1 if all elements in the array are different and 0 otherwise. The algorithm runs in 2 steps: 1) $\mathcal{P}$ provides an ascending sorted array $A^{\ast}$ that is a permutation of array $A$ and 2) $\mathcal{V}$ checks that all elements in $A^{\ast}$ are different. Checking 2) is trivial as $A^{\ast}$ is sorted. To solve 1), $\mathcal{V}$ hashes both $A$ and $A^{\ast}$ using an associative hash function $\mathcal{AH}$ and checks if these 2 hash values are the same.

\begin{algorithm}
  \caption{Verify all elements in an array are different}\label{alg:element_uniqueness}

  \textbf{Input}: unsorted array $A$, ascending sorted array $A^{\ast}$, associative commitment function $\mathcal{AH}$\\
  \textbf{Output}: 1 if all elements in $A$ are different and 0 otherwise.

  \begin{algorithmic}
      \Function{PairwiseDistinctCheck}{$A, A^{\ast}, \mathcal{AH}$}
          \State {$h1$ $\gets$ {$\mathcal{AH}(A)$}}
          \State {$h2$ $\gets$ {$\mathcal{AH}(A^{\ast})$}}

          \If{$h1\neq h2$}
              \State \Return {$0$}
          \EndIf

          \State {$n$ $\gets$ {$A^{\ast}$.len()}}

          \For{$i \gets 1$ to $(n - 1)$}
              \If{$A^{\ast}_i\leq A^{\ast}_{i-1}$}
                  \State \Return {$0$}
              \EndIf
          \EndFor

          \State \Return {$1$}
      \EndFunction
  \end{algorithmic}
\end{algorithm}

We use the algorithm \ref{alg:element_uniqueness} to show that all public keys are different. To prove that validators are distinct across different blocks, $\mathcal{P}$ needs to include all sorted public key indexes from the first block of the epoch to the current block as part of the witness. If this is the correct list of indexes, it's easy to check if all elements are distinct using the algorithm above. There could be at most one million indexes in an epoch but they could be divided into \(M\) subcircuits and each subcircuit would handle a reasonable amount of indexes.

In order to make sure that $\mathcal{P}$ provides a correct list of indexes, we include one more number $\mathcal{H}_{pre}$ in the input of our proof besides current block hash and previous beacon state hash. $\mathcal{H}_{pre}$ is the result of applying $\mathcal{AH}$ to all public key indexes from block $0$ to block ${(i-1)}$ assuming that we are processing block $i^{th}$. $\mathcal{V}$ uses $\mathcal{H}_{pre}$ as input and continues to hash it with other public key indexes in current block to create a new value $\mathcal{H}_{cur}$. Note that since $\mathcal{AH}$ is an associative hash the value $\mathcal{H}_{cur}$ is equivalent to applying $\mathcal{AH}$ to all public key indexes from block $0$ to block $i^{th}$. Finally, $\mathcal{V}$ compares $\mathcal{H}_{cur}$ with the hash of index list provided by $\mathcal{P}$ to make sure the list is correct.

The value of $\mathcal{H}_{pre}$ is kept on-chain in the smart contract. The smart contract uses this value as one of the inputs for the Groth16 verification step. Upon successful verification, the contract updates this $\mathcal{H}_{pre}$ with the $\mathcal{H}_{cur}$ to be used in later blocks.

\subsubsection{Accumulation gate}
Accumulation gate is a special type of gate that contains a list of nested gates, each of which is a fan-in 2 gate. The value of an accumulated gate is the sum of its nested gate. This allows multiple input gates to connect to the same output gate. However, an accumulation gate is not a fan-in \(n\) gate as it does not directly connect to \(n\) other input gates. Instead, it could be thought of as a container of multiple fan-in 2 gates which might be of different types and come from different layers.

Recall our formula for calculating MLE of output gate $g$ in layered circuit is as follow:
\[ \tilde{V_i}(g) = \sum_{x,y \in \{0,1\}^{2s_{i+1}}} \widetilde{add_i}(g,x,y)(V_{i+1}(x)+V_{i+1}(y))+\widetilde{mult_i}(g,x,y)(V_{i+1}(x).V_{i+1}(y))\]
$\tilde{V_i}(g)$ is in fact the sum of all the input gates that connect to the same output gate. In other words, for the same output $g$ we can have multiple input tuples $(x,y)$ where $(x,y)$ are input gates that connect to.

\begin{figure}[H]
    \centering
    \includegraphics[scale=0.5]{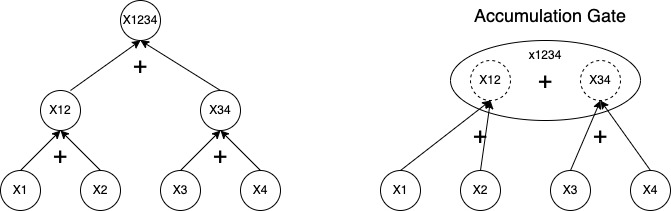}
    \caption{An example circuit that adds 4 numbers without and with an accumulation gate.}
    \label{fig:accumulation-gate}
\end{figure}

Figure \ref{fig:accumulation-gate} demonstrates how to add 4 numbers without and with accumulation gate. Without an accumulation gate, the left circuit needs 3 gates, 6 wirings on 2 layers. With accumulation, we only need 1 gate, 4 wiring and 1 layer. Note that the nested gates inside the accumulation gate must be fan-in 2 gates though they can be of different types. Accumulation gates can reduce the number of layers and gates several times. In practice, our SHA256 circuit uses only 4 layers compared to 14 layers without an accumulation gate.

\section{Implementation and Evaluation}
We implement a prototype of Sisu protocol with focus on the heavy computation part. We first implement our protocol to run on CPU with Rust and then gradually port the code to run on GPU. Our goal is to run both phase 1 and phase 2 under one ethereum block time frame. The first phase contains both Rust and C++ code (20000+ lines of code) and the second part is implemented in Circom with many parts of the code being generated. The code is open source for community feedback and improvement \autocite{sisubridge}.

\subsection{Implementation Details}
The following part is implemented for evaluation at the time of writing:
\begin{itemize}
  \item Public key aggregation for a set of nodes. One public key addition requires 32,000 input gates.
  \item SHA256 circuit for hash function used in Ethereum consensus. Each circuit has 8096 inputs with approximately 50,000 gates.
  \item Merkle path proof of validator in BeaconState. The size of a Merkle path is 50 but with optimization we only need to generate proof for a path with size 16. Each node requires 2 SHA256 hashes and the total SHA hashes in one Merkle path is 32. Proving one Merkle path requires 1.6M circuit gates and 256K input.
\end{itemize}

The dominating factor in prover's time spent on proving the Merkle path since we have to generate proof for every validator. At 32,000 validators, the total circuit size for Merkle path proof is 51,2B gates. Our prototype does not include distinct public key proof but we expect this is minor compared to the size of validator's Merkle path proof. This implementation is not fully optimized and there is much room for future improvements.

\textbf{Choice of Finite Field}. There are 2 finite fields under our consideration: Goldilocks ($2^{64} - 2^{32} + 1$) and BN254. Goldilocks is fast for provers (4-5x faster than BN254) but does not support pairing. If we want to use Goldilocks, we have to emulate the operation of Goldilocks in another field that supports pairing. This significantly increases the number of constraints in Groth16. For that reason, our implementation uses BN254 for better verifier performance.

\subsection{Evaluation}

\subsubsection{Hardware setup}
We use machines with Intel(R) 8 cores i7-7700 @3.60GHz CPU, 32GB RAM and Geforce GTX 1080 (8GB VRAM) GPU. The cost of one Geforce GTX 1080 GPU is about 230 USD and significantly cheaper than other modern GPU. We use modest hardware configuration to show that our implementation can run efficiently with average hardware. Moreover, as our protocol does not have work-to-worker communication our implementation can scale to hundred or even thousands of machines with low overhead. The cheap hardware configuration makes the overall system competitive with other proof systems in terms of price. The ability to easily scale opens an opportunity to extend Sisu protocol to become a distributed proof generation network.

\subsubsection{Phase 1 Evaluation}
\textbf{Single GPU}. We run experiments for 1, 2, 4, 8, 16 validators on a single GPU. In each experiment, there is a base task running in 6s for VPD commitment and running FFT GKR regardless of the number of validators. When the number of validators doubles we observe a small increment in the running time.

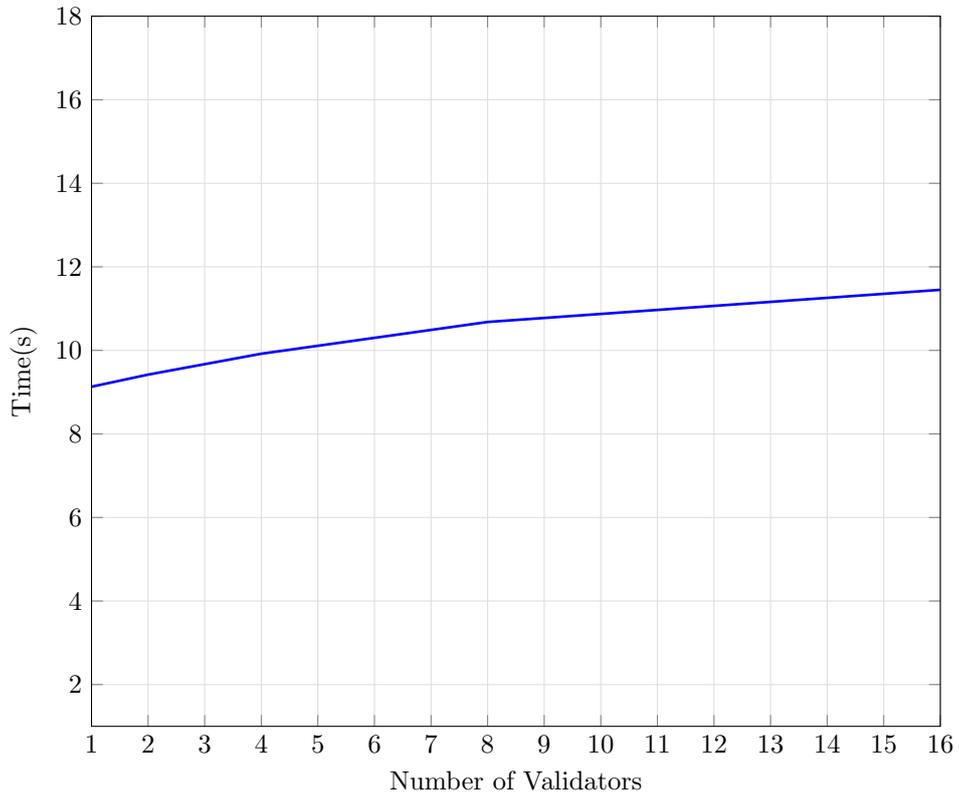
\begin{figure}[H]
  \centering
  \begin{tikzpicture}
    \begin{axis}[
      xlabel=Number of Validators,
      ylabel=Time(s),
      grid=both,
      xtick={1,2,...,16},
      xmin=1,
      ymin=1,
      xmax=16,
      ymax=18,
      minor grid style={gray!25},
      major grid style={gray!25},
      width=0.75\linewidth,
      no marks]
    \addplot[line width=1pt,solid,color=blue] %
      table[x=numval,y=time,col sep=comma]{eval-single-gpu.csv};
    \end{axis}
  \end{tikzpicture}

  \caption{Running time with single GPU}
  \label{fig:single-gpu}
\end{figure}

\textbf{Multi GPUs}. We experiment with the performance of the system when running a cluster with multiple GPUs. Compared to single GPU execution, there is additional time spent on networking and moving data between GPU and CPU. We see a moderate increase in running time between 1 and 2 GPU but there is little difference in running time between 2 and 4 GPUs.

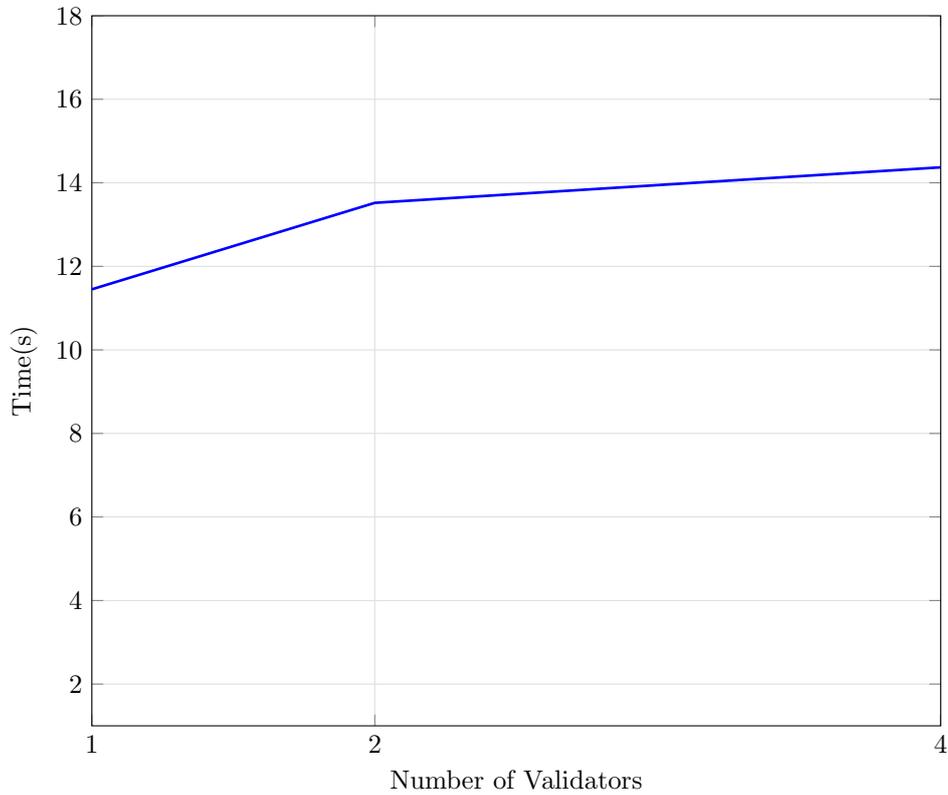
\begin{figure}[H]
  \centering
  \begin{tikzpicture}
    \begin{axis}[
      xlabel=Number of Validators,
      ylabel=Time(s),
      grid=both,
      xtick={1,2,4},
      xmin=1,
      ymin=1,
      xmax=4,
      ymax=18,
      minor grid style={gray!25},
      major grid style={gray!25},
      width=0.75\linewidth,
      no marks]
    \addplot[line width=1pt,solid,color=blue] %
      table[x=numval,y=time,col sep=comma]{eval-multi-gpu.csv};
    \end{axis}
  \end{tikzpicture}

  \caption{Running time with multiple GPUs}
  \label{fig:multi-gpu}
\end{figure}

\subsubsection{Phase 2 Evaluation}
The circuit in Phase 2 emulates a verifier program in Phase 1. The circuit of phase 2 contains 2 major parts: one doing signature verification against a public key and the other verifies Merkle path in polynomial commitments. The signature verification is done once regardless of the number of validators. The second part is dependent on the cluster size. We take the implementation from \autocite{circompairing} for BLS12-381 signature verification which has 19.2M constraints.

If we choose cluster size to be 128 validators (i.e. 8 GeForce 1080 GPUs), we need 4 clusters to verify 512 validators. The number of constraints is 29.45M. To verify 32,000 validators, we need 128 clusters with 339.45M constraints. The number of constraints could be reduced by increasing the cluster size. For example, if the cluster size is 512, the number of constraints reduces to 179.45M. With cluster size equal 1024, we only need 99.45M constraints to verify all Merkle paths.

The Groth16 running time using Circom is dominated by 2 factors: witness generation time and proof generation time. Proof generation time consists of mostly multi scalar multiplication (MSM). MSM operations can run in parallel. One R1CS in Groth16 requires 3 MSMs on G1 and 1 MSM on G2 or about 5-6 MSMs on G1 since operation on G2 is 2-3x slower than in G1. In our case we can save one MSM operation since we do not have to hide the witness. Therefore the number of MSM operations is about 4-5x the number of constraints. Though the total of MSMs can be large, their execution time could be below 12s if we have enough GPUs. Performance of MSMs operation on a single GeForce 1080 GPU is as below.

\begin{center}
  \begin{tabular}{|l|l|l|l|}
    \hline
\thead{No. MSMs} & \thead{Running time} \\ \hline
$2^{19}$ & 266ms \\    \hline
$2^{22}$ & 943ms \\    \hline
$2^{24}$ & 3577ms \\   \hline
    \end{tabular}
\end{center}

Unlike proof generation, the witness generation cannot run in parallel. A circuit with 25M R1CS constraints takes 2.5 minutes to generate a witness. We are using rapidsnark \autocite{rapidsnark} for witness generation. Optimizing rapidsnark is beyond the scope of this paper and could be reserved for future work.

Another direction we can consider is replacing Groth16 with Halo2 libraries with advanced lookup techniques like Lasso \autocite{cryptoeprint:2023/1216}. Halo2 could provide faster prover time compared to Groth16 but requires considerable expertise and manual optimization to fully utilize the framework.

\section{Conclusion}

Building a trustless bridge for Ethereum full node is a challenging task as it requires huge computation resources. We propose a new protocol Sisu with emphasis on decentralized proving. The development and emergence of prover markets will make distributed proof a viable option in the future. We implement an open source prototype to demonstrate our concept as well as highlight the remaining tasks.

Sisu protocol could be modified to support other types of ambitious ZK applications and not just a bridge. We expect a fully optimized version could support tens of billions of gates and half a billion of R1CS constraints running in distributed manner. We make our code open source for the community to advance the development of trustless ZK applications including trustless bridges.

\printbibliography

\end{document}